\documentclass[aps,showpacs,showkeys,superscriptaddress,nofootinbib,preprint,11pt]{revtex4}
\usepackage{graphicx}% Include figure files
\usepackage{subfigure}

\usepackage{graphicx}% Include figure files
\def\slashchar#1{\setbox0=\hbox{$#1$}
   \dimen0=\wd0 \setbox1=\hbox{/} \dimen1=\wd1
   \ifdim\dimen0>\dimen1 \rlap{\hbox to \dimen0{\hfil/\hfil}} #1
   \else  \rlap{\hbox to \dimen1{\hfil$#1$\hfil}} / \fi}

\begin{document}

\title{Weak $\eta$ production off the nucleon}

\author{M. Rafi Alam}
\affiliation{Department of Physics, Aligarh Muslim University, Aligarh-202 002, India}

\author{L. Alvarez-Ruso}
\affiliation{Departamento de F\'\i sica Te\'orica and Instituto de F\'isica Corpuscular, Centro Mixto
Universidad de Valencia-CSIC, E-46071 Valencia, Spain}

\author{M. Sajjad Athar}
\affiliation{Department of Physics, Aligarh Muslim University, Aligarh-202 002, India}

\author{M. J. Vicente Vacas}
\affiliation{Departamento de F\'\i sica Te\'orica and Instituto de F\'isica Corpuscular, Centro Mixto
Universidad de Valencia-CSIC, E-46071 Valencia, Spain}

\begin{abstract}
The weak $\eta$-meson production off the nucleon induced by (anti)neutrinos is studied at 
low and intermediate energies, the range of interest for several ongoing and future neutrino experiments. We consider Born diagrams and the excitation of $N^\ast (1535)S_{11}$ and $N^\ast(1650)S_{11}$ resonances. The vector part of the N-$S_{11}$ transition form factors has been obtained from the MAID helicity amplitudes while the poorly known axial part is constrained with the help of the partial conservation of the axial current (PCAC) and assuming the pion-pole dominance of the pseudoscalar form factor. 
\end{abstract}
\maketitle
%%%%%%%%%%%%%%%%%%%%%%%%%%%%%%%%%%%%%%%%%%%%
%% MAINMATTER
%%%%%%%%%%%%%%%%%%%%%%%%%%%%%%%%%%%%%%%%%%%%
\section{Introduction}
Eta production through electromagnetic interactions has been extensively 
 studied both theoretically and experimentally. 
However, for $\eta$ production induced by weak interactions there are no results neither
for the integrated cross section nor for distributions.
Its study is interesting because of several reasons. For example, 
$\eta$ is one of the important probes to search for the strange quark content of the nucleons~\cite{Dover:1990ic}.
Also a precise determination of the $\eta$ production cross section would help in
subtracting the background in proton decay searches. In some supersymmetric grand unified theories,  
$\eta$ mesons provide a prominent signal for proton decay~\cite{Wall:2000pq}. Therefore, its background contribution 
due to atmospheric neutrino interactions should be well estimated. 
Furthermore, the $\eta$ production channel is likely to be dominated by $N^\ast(1535)$ resonance excitation. 
This state sits near the threshold of the $N \eta$ system and has large branching ratio into the $N \eta$  channel.
Thus, a precise measurements of the cross section will also allow to determine the axial properties of this resonance. 
In addition, a second class (via pion pole) $\eta$ production mechanism that could compete with resonance production in certain kinematic conditions has been singled out~\cite{Dombey:1968vh}. Finally, theoretical models like the present one allow to improve the Monte Carlo simulations used to analyze oscillation experiments. 

We have studied charged current (CC) $\eta$ production off the nucleon induced by (anti)neutrinos.
Born terms are calculated using a microscopical model based on the SU(3) chiral Lagrangian. 
The basic parameters  are $f_\pi$, the pion decay constant, Cabibbo's angle, 
the proton and neutron magnetic moments and the axial vector coupling constants for the baryon octet, $D$ and $F$,
 that are obtained from the analysis of the semileptonic decays of neutron and hyperons.
 We also consider $N^\ast$(1535) and $N^\ast$(1650) $S_{11}$ resonant intermediate states.
 The vector form factors of the N-$S_{11}$ transition have been obtained 
 from the helicity amplitudes extracted in the analysis of world pion photo- and electroproduction data with the unitary isobar model MAID~\cite{Drechsel:2007if}. The properties of the axial N-$S_{11}$ transition current are basically unknown but assuming the pion-pole dominance of the pseudoscalar form factor, together with PCAC one can fix the axial coupling using the empirical $N^* \rightarrow N \pi$ partial decay width. We make an educated guess for the dependence 
on the 4-momentum squared transferred by the neutrino, which ultimately remains to be determined experimentally.   

The paper is organized as follows. First we present the formalism for CC $\eta$ production 
in  neutrino/antineutrino nucleon scattering  based on the Lagrangian of SU(3) $\chi$PT and describe the 
input for the $S_{11}$ resonances. Finally we discuss the results and summarize our study. 

\section{Formalism}\label{Formalism}

The reactions for the charged current (anti)neutrino induced eta production on free nucleons are
\begin{eqnarray}\label{reaction}
\nu_{l}(k) + n(p) \rightarrow l^-(k^{\prime}) + p(p^{\prime}) + \eta(p_{2}) \; ,  \qquad
\bar \nu_{l}(k) + p(p) \rightarrow l^+(k^{\prime}) + n(p^{\prime}) + \eta(p_{2}).
\end{eqnarray}
The amplitudes for these processes are given in terms of the product of the leptonic and hadronic currents. 
The standard V-A current is taken for the leptonic part while for the hadronic current we consider tree level Born 
diagrams (s- and u-channels) with intermediate nucleon and $S_{11}$ resonances. The non-resonant terms are obtained using the SU(3) chiral Lagrangian ~\cite{Rafi:2010} at leading order. Other contributions from this Lagrangian are ruled out by the various symmetry arguments.  We get
\begin{eqnarray}\label{Eq:amp}
{\cal J}_{N(s)}^\mu &=&  \frac{g V_{ud} }{2\sqrt2} \frac{D-3F}{2\sqrt3 f_\pi} \bar u_N (p^\prime) \slashchar{p_2} \gamma^5  \frac{\slashchar{p}+\slashchar{q}+M}{(p+q)^2-M^2} 
\left( \gamma^\mu - (D+F) \gamma^\mu  \gamma^5 \right)u_N (p) \nonumber \\ 
{\cal J}_{N(u)}^\mu &=&  \frac{g V_{ud} }{2\sqrt2} \frac{D-3F}{2\sqrt3 f_\pi} \bar u_N (p^\prime) \left( \gamma^\mu - (D+F) \gamma^\mu  \gamma^5 \right)
  \frac{\slashchar{p}-\slashchar{p_2}+M}{(p - p_2)^2-M^2} 
\slashchar{p_2} \gamma^5 u_N (p) \nonumber \\
{\cal J}_{R(s)}^\mu & = & \frac{g V_{ud} }{2\sqrt2} i g_{\eta N S_{11}} \bar{u}_N (p^\prime)  
 \frac{ \slashchar{p} +\slashchar{q} + M_R }{(p+q)^2 - M_R^2 + i \Gamma_R M_R} {\cal O}^\mu u_N (p) \nonumber \\
{\cal J}_{R(u)}^\mu & = &  \frac{g V_{ud} }{2\sqrt2} i g_{\eta N S_{11}} \bar{u}_N (p^\prime) {\cal O}^\mu
 \frac{ \slashchar{p} -\slashchar{p}_2 + M_R }{(p - p_2)^2 - M_R^2 + i \Gamma_R M_R} u_N (p) \nonumber \\
{\cal O}^\mu &=&\frac{F_1^{V}(Q^2)}{(2 M)^2}(Q^2\gamma^\mu + \slashchar{q} q^\mu) \gamma_5
\pm \frac{F_2^{V}(Q^2)}{2 M} i \sigma^{\mu\rho} q_\rho \gamma_5
- F_A (Q^2) \gamma^\mu  \mp \frac{F_P (Q^2)}{M} q^\mu
\end{eqnarray}
where the subscripts $(s)$ and $(u)$ denote the s- and u-channel diagrams.
The upper (lower) sign in ${\cal O}^\mu$ applies to the s-(u-)channel current. 
$M_R$ and $\Gamma_R$ are the mass and total decay width of the $S_{11}$ resonances while $M$ denotes the nucleon mass. The momentum transfer is denoted by $q=k-k'$ while $Q^2 = -q^2$; 
$g$ is the gauge coupling, which is related to the Fermi coupling constant $G_F=\sqrt{2} g^2/(8 M^2_W)$. The isovector form factors $F_{1,2}^{V}$, are given in terms of the electromagnetic transition form factors of protons and neutrons as
\begin{equation}
 F_1^V(Q^2) = F_1^ p(Q^2) - F_1^n(Q^2) ;\quad F_2^V(Q^2)= F_2^p(Q^2) - F_2^n(Q^2) .
\end{equation}
$F_{1,2}^{p,n}(Q^2)$ can then be obtained from the helicity amplitudes $A_{\frac{1}{2}}^{p,n}$, and $S_{\frac{1}{2}}^{p,n}$, which have been conveniently parametrized in Ref.~\cite{Drechsel:2007if}. In our case~\cite{Leitner:2008ue}
\begin{eqnarray*}
A_{\frac{1}{2}}^{p,n} &=& \sqrt{\frac{2 \pi \alpha_e }{M}\frac{(M_R+M)^2+Q^2}{M_R^2-M^2}} 
\left( \frac{Q^2}{4M^2} F_1^{p,n}(Q^2) + \frac{M_R-M}{2M} F_2^{p,n}(Q^2) \right) \nonumber \\
S_{\frac{1}{2}}^{p,n} &=& \sqrt{\frac{\pi \alpha_e }{M}\frac{(M_R - M)^2+Q^2}{M_R^2-M^2}} 
\frac{(M_R + M)^2+Q^2}{4 M_R M}
\left(  \frac{M_R-M}{2M} F_1^{p,n}(Q^2) - F_2^{p,n}(Q^2) \right)
\end{eqnarray*}
where $\alpha_e$ is the fine structure constant. The quality of the model and the sensitivity to the parameters can be tested against photo and electroproduction data\footnote{Paper in preparation.}. 

For the axial form factor $F_A ( Q^2 )$ we have adopted a dipole form with  $M_A =1.05$~GeV. The pseudoscalar form factor is related  to $F_A ( Q^2 )$ through the PCAC relation
\begin{equation}
F_A ( Q^2 ) = F_A (0) \left( 1 + \frac{Q^2}{M_A^2} \right)^{-1}; \qquad
F_P ( Q^2 ) = \frac{( M_R - M ) M}{Q^2 + m_\pi^2} F_A ( Q^2 ).
\end{equation}
In the numerical calculations we have used $F_A(0)=0.2121$ for $S_{11}(1535)$ and $F_A(0)=0.2105$ for $S_{11}(1650)$ from the corresponding off-diagonal Goldberger-Treiman relations. The coupling $g_{\eta N S_{11}}$ is obtained from the $S_{11} \rightarrow N \eta$ decay width calculated using the Lagrangian
\begin{equation}\label{Eq:etaNS11Lag}
{\cal L}_{\eta NS_{11}} = -i g_{\eta N S_{11}} \bar\Psi \Psi_{S_{11}} \phi_{\eta}  + \textrm{h.c.},
\end{equation}
where $\Psi$ and $\Psi_{S_{11}}$ are the Dirac fields for the nucleon and $S_{11}$ resonance respectively, while $\phi_{\eta}$ is the $\eta$ meson field. The total decay width for these two resonances is given by,
\begin{eqnarray}
 \Gamma_R (1535) &=& 0.42 \; \Gamma_{N^\ast \rightarrow N \eta } \; +  0.46 \; \Gamma_{N^\ast \rightarrow N \pi } \;
+ 0.12 \; \Gamma_{N^\ast \rightarrow NX } \nonumber \\
\Gamma_R (1650) &=& 0.10 \; \Gamma_{N^\ast \rightarrow N \eta } \; +  0.7 \; \Gamma_{N^\ast \rightarrow N \pi } \;
+ 0.2 \; \Gamma_{N^\ast \rightarrow NX }. \nonumber
\end{eqnarray} 
Here we would like to emphasize that there are large uncertainties in the parameters of the $S_{11}$ resonances including a large error bar in their decay widths. 

\section{Results and Discussion}\label{Results}

\begin{figure}
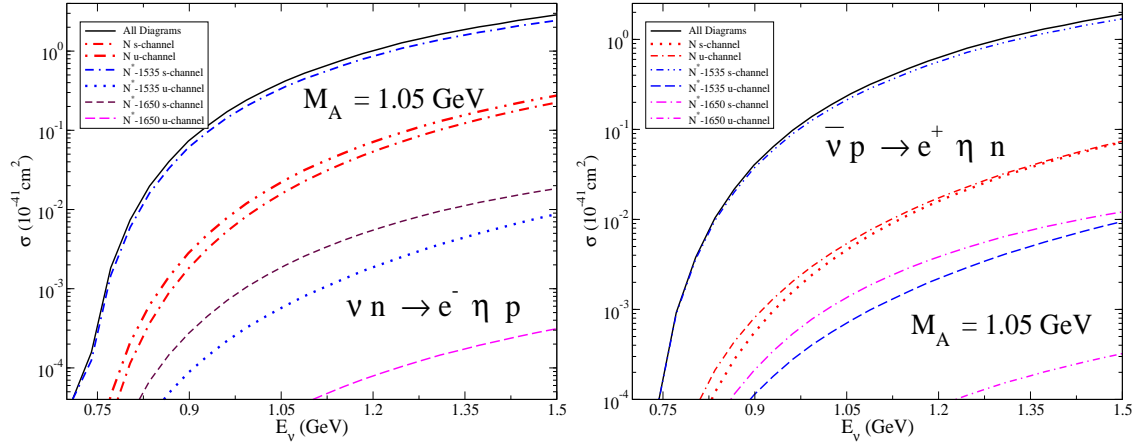

\includegraphics[height=.25\textheight,width=0.45\textwidth]{nu.eps}
\includegraphics[height=.25\textheight,width=0.45\textwidth]{nubar.eps}
\caption{Contribution of the different terms to the total cross section for the  $ \nu_{e} + n \rightarrow e^- + p + \eta$ reaction (left panel) and
 $ \bar\nu_{e} + p \rightarrow e^+ + n + \eta$ reaction (right panel).}
\label{fig:xsec}
\end{figure}

The cross sections ($\sigma$) corresponding to the amplitudes given in Eq.~\ref{Eq:amp} are shown in Fig.~\ref{fig:xsec} for $\nu/\bar\nu$ energy up to $1.5$~GeV. We find that the $N^\ast (1535)$ resonance is dominant while the contribution of $N^\ast(1650)$ to the total cross section is small. 
This can be understood easily because $N^\ast(1535)$ is lighter and has a relatively larger branching ratio into  $\eta N$ than $N^\ast(1650)$. The contribution of the non-resonant diagrams in case of neutrino induced CC process is higher than in the corresponding  anti neutrino channels. Also in neutrino mode the contribution of u-channel diagram is slightly larger than the corresponding s-channel diagram. Furthermore, we have studied $Q^2$, lepton energy and angular distributions for these processes and the results would be communicated elsewhere. 

To conclude, in this work we have studied the CC $\nu/\bar\nu$ induced $\eta$ production on free nucleons. We find that the cross sections are large enough to be measured in experiments like T2K, NO$\nu$A and MINER$\nu$A.

\section{ACKNOWLEDGMENTS}

MRA and MSA are thankful to the Aligarh Muslim University for the financial support to attend NuInt12. This work is partially  supported by the Spanish Ministerio de Econom\'ia y Competitividad and European FEDER funds under Contracts FIS2011-28853-C02-01 and  FIS2011-28853-C02-02, Generalitat Valenciana under Contract PROMETEO/2009/0090 and the EU Hadron-Physics2 project, Grant No. 227431.

\end{document}